\documentclass[letterpaper,twocolumn,10pt]{article}
\usepackage{usenix2019_v3}
\usepackage{graphicx}
\usepackage{booktabs}
\usepackage{listings}
\usepackage{amsmath}
\usepackage{centernot}
\usepackage{xspace}
\usepackage{enumitem}
\usepackage{array}
\usepackage{xurl}
\usepackage{tabularx}
\usepackage{ragged2e}
\usepackage{caption}
\hypersetup{breaklinks=true,linkcolor=black,citecolor=black,urlcolor=black}
\newcolumntype{Y}{>{\RaggedRight\arraybackslash}X}
\usepackage{tikz}
\usetikzlibrary{arrows.meta,positioning,fit,calc}
\newcommand{\avgidle}{\ifmmode\mathtt{avg\_idle}\else\texttt{avg\_\allowbreak idle}\fi\xspace}
\newcommand{\idlestamp}{\ifmmode\mathtt{idle\_stamp}\else\texttt{idle\_\allowbreak stamp}\fi\xspace}
\newcommand{\newidle}{\ifmmode\mathtt{sched\_balance\_newidle()}\else\texttt{sched\_\allowbreak balance\_\allowbreak newidle()}\fi\xspace}
\newcommand{\putprevidle}{\ifmmode\mathtt{put\_prev\_task\_idle()}\else\texttt{put\_\allowbreak prev\_\allowbreak task\_\allowbreak idle()}\fi\xspace}
\newcommand{\updaterq}{\ifmmode\mathtt{update\_rq\_avg\_idle()}\else\texttt{update\_\allowbreak rq\_\allowbreak avg\_\allowbreak idle()}\fi\xspace}
\newcommand{\ttwupending}{\ifmmode\mathtt{ttwu\_pending}\else\texttt{ttwu\_\allowbreak pending}\fi\xspace}
\newcommand{\maxidlecost}{\ifmmode\mathtt{max\_idle\_balance\_cost}\else\texttt{max\_\allowbreak idle\_\allowbreak balance\_\allowbreak cost}\fi\xspace}
\newcommand{\utilest}{\ifmmode\mathtt{util\_est}\else\texttt{util\_\allowbreak est}\fi\xspace}
\newcommand{\peltidleclock}{\ifmmode\mathtt{update\_idle\_rq\_clock\_pelt()}\else\texttt{update\_\allowbreak idle\_\allowbreak rq\_\allowbreak clock\_\allowbreak pelt()}\fi\xspace}
\newcommand{\picknextfair}{\ifmmode\mathtt{pick\_next\_task\_fair()}\else\texttt{pick\_\allowbreak next\_\allowbreak task\_\allowbreak fair()}\fi\xspace}
\newcommand{\setnextidle}{\ifmmode\mathtt{set\_next\_task\_idle()}\else\texttt{set\_\allowbreak next\_\allowbreak task\_\allowbreak idle()}\fi\xspace}
\newcommand{\utilestenqueue}{\ifmmode\mathtt{util\_est\_enqueue()}\else\texttt{util\_\allowbreak est\_\allowbreak enqueue()}\fi\xspace}
\newcommand{\updateloadavg}{\ifmmode\mathtt{update\_load\_avg()}\else\texttt{update\_\allowbreak load\_\allowbreak avg()}\fi\xspace}
\newcommand{\enqueuetaskfair}{\ifmmode\mathtt{enqueue\_task\_fair()}\else\texttt{enqueue\_\allowbreak task\_\allowbreak fair()}\fi\xspace}
\newcommand{\schedutil}{\texttt{schedutil}\xspace}

\begin{document}
\raggedbottom
\date{}
\hypersetup{pdftitle={Resolving an Idle-Time Accounting Hazard in Linux Load Balancing},pdfauthor={Shubhang Kaushik}}
\title{\Large\bf Resolving an Idle-Time Accounting Hazard in Linux Load Balancing}
\author{{\rm Shubhang Kaushik}\\Ampere Computing\\\texttt{shubhangpk@gmail.com}}
\maketitle

\begin{abstract}
The Linux kernel scheduler uses per-runqueue idle-time history
(\avgidle) to decide whether searching other CPUs for runnable work
justifies its execution overhead. This estimate is an active input to
new-idle load balancing, helping determine whether a search continues
across scheduling domains.

This operational study presents the design and subsequent hardening of
idle-time accounting through two upstream patch series. The original
wakeup-side accounting missed legitimate idle exits associated with task
creation, exec-related placement paths, and load-balancing migration.
Introducing \updaterq and moving accounting to idle-task switch-out
broadened coverage beyond wakeup-specific arrival paths. However, the
broader consumer could also reach histories without a valid start
timestamp. Subtracting zero from the runqueue clock constructs a sample
unrelated to the measured idle interval and can rapidly saturate the
bounded estimator, weakening new-idle balance cost gating.

We describe the accounting redesign and its refinement using public
development archives, patch iterations (v1--v8 and v1--v3), runtime tracing,
independent review, and upstream integration records. A
configuration-independent \ttwupending history reaches the relocated
consumer without a compatible measurement start. Core Scheduling
force-idling and proxy execution further show why mechanically symmetric
idle-entry timestamping is insufficient. From this incident and two
retrospective scheduler cases, we extract a state-lifetime audit organized
around missing establishment, premature consumption, and stale
consumption. We distinguish these correctness findings from performance
attribution, for which a controlled three-variant experiment remains necessary.
\end{abstract}

\section{Introduction}\label{sec:intro}
Operating-system schedulers rely on state accumulated across execution
events to guide resource allocation. In Linux, per-runqueue idle-time
history participates in the cost decisions governing new-idle load
balancing across multicore scheduling domains. Once historical state
feeds such policy, its \emph{semantic lifetime}\ becomes part of policy
correctness.

A consumer can observe semantically invalid state even when the
arithmetic is locally sound. An expected producer may not have executed
along a control path; a consumer may run before prerequisite updates; or
an initialized value from an earlier epoch may survive into a context
where its meaning is invalid.

This paper investigates that boundary through the introduction of
\updaterq, the relocation of idle-time accounting to \putprevidle,
and the subsequent refinement of the helper's measurement-validity check.
The redesign changes the event at which idle-duration state is consumed;
the follow-up repair refines the validity condition at that broader boundary.

The original accounting consumed \idlestamp along wakeup-oriented
paths. It failed to capture some transitions where idle CPUs acquired
runnable work through task creation, exec-related placement, or
load-balancing migration \cite{idle_coverage}. Consequently, some idle
exits occurred without contributing a sample to \avgidle. The resulting
bias depends on the omitted intervals and the existing estimate; missing
samples alone do not establish that the estimate is always too low.

The upstream change extracted \updaterq and relocated consumption to
\putprevidle, binding accounting to switching away from the idle task
rather than to how work arrived \cite{idle_relocation,idle_relocation_merge}.
Testing reported with the patch on an 80-core Ampere Altra using a
6.19-rc5 baseline reported a 7.2\% hackbench performance improvement at
16 threads and reduced schbench $p99.9$ latency at high concurrency
\cite{idle_relocation}. These are configuration-specific patch-submission
results, not new measurements performed for this manuscript.

Independent evaluation by K.~Prateek Nayak at AMD found substantial
changes in scheduler activity: SMT-domain new-idle attempts rose from
3,964 to 17,961, and MC-domain attempts from 1,917 to 7,022. The same
evaluation repeatedly observed sensitivity at tbench-256, while noting
that profiling reruns reduced the apparent throughput difference to
approximately 2\%, within the stated margin of error
\cite{idle_relocation_review}.

The relocation also removed the nonzero-\idlestamp check present at the
old consumer. Masahito S. independently observed that subtracting a zero
timestamp constructs a runqueue-clock sample rather than elapsed idle
time and can drive \avgidle to its $2\times\maxidlecost$ clamp
\cite{idle_symmetric_proposal}. Temporary tracing under hackbench later
confirmed that \updaterq can execute with \texttt{idle\_stamp == 0}
\cite{idle_guard_v3}.

Upstream review identified a configuration-independent reachable history:
when \ttwupending is set, \newidle returns before establishing
\idlestamp. The runqueue can subsequently execute the idle task and
switch away from it after processing the pending wakeup
\cite{idle_source_snapshot,idle_saturation_review}.

A symmetric-entry proposal would timestamp from \setnextidle. But
idle-task selection can occur under force-idling or proxy execution
without representing the beginning of the ordinary interval summarized
by \avgidle \cite{idle_symmetric_review,idle_guard_v3}. Thus:
\begin{equation}\label{eq:symmetry}
\begin{aligned}
\text{control-flow symmetry}\\
\centernot{\Rightarrow}\text{measurement-lifetime symmetry}.
\end{aligned}
\end{equation}
The eventual correction retained the switch-out consumer and restored the
nonzero guard \cite{idle_guard_merge}. It preserves the coverage change
while rejecting the demonstrated histories lacking measurement establishment.

\paragraph{Contributions.}
First, the paper presents the helper introduction, accounting relocation,
and subsequent validity refinement, supported by patch revisions,
review, runtime observation, and upstream integration. Second, it
separates semantic coverage from producer compatibility when a consumer
moves. Third, it extracts a practical review procedure and applies it
retrospectively to independent PELT and \utilest/\schedutil failures.
The historical record does not isolate performance effects across all
three accounting variants; that limitation remains explicit.

\section{Operational Model and Review Questions}\label{sec:model}
We model a stateful policy input as
\begin{equation}\mathcal{M}=(S,C,G),\end{equation}
where $S$ establishes or updates state, $C$ consumes it for a
policy-relevant operation, and $G$ is the implementation predicate
admitting consumption. The consumer operation is distinct from its
syntactic site $E_C$. Each history considered here is truncated at one
particular occurrence of that site; ordering is relative to that occurrence,
not to a static source location across repeated executions. Let $R(E_C)$
be these histories reaching the site.
The implementation admits consumption on
\begin{equation}\label{eq:implementation-consumption}
R_G(C)=\{h\in R(E_C)\mid G(h,E_C)\}.
\end{equation}
For each history $h$, let $S_h$ denote a relevant establishment event.
Define semantically compatible histories as
\begin{equation}\label{eq:compatible-establishment}
\begin{aligned}
R_S(E_C)=\big\{h\in R(E_C)\;\big|\;&\exists S_h:S_h\prec_h E_C\\
&\land\operatorname{Compat}(S_h,E_C,h)\big\}.
\end{aligned}
\end{equation}
For the examined cases, compatibility requires checking at least:
\begin{enumerate}[leftmargin=*,nosep]
\item \textbf{Semantic identity:} Does the establishment represent the
intended measurement interval or execution event?
\item \textbf{Ordering:} Does it follow prerequisite transitions and
precede consumption?
\item \textbf{Freshness:} Does the state remain current for this consumption,
rather than represent a superseded execution state?
\end{enumerate}
The review obligation is
\begin{equation}\label{eq:safety}R_G(C)\subseteq R_S(E_C).\end{equation}
This is a semantic audit criterion, not an automated verifier. The
nonzero timestamp guard rejects the demonstrated missing-establishment
histories; nonzero alone does not generally prove provenance, ordering,
or freshness.

\subsection{Three Observed Lifetime Hazards}\label{sec:hazards}
\paragraph{H1: Missing measurement establishment.}
For an admitted consumer occurrence, no producer established a measurement
for the intended epoch. This is narrower than the general incompatibility
condition $h\in R_G(C)\land h\notin R_S(E_C)$, which can also result from
incorrect ordering or stale state. In this case, the zero sentinel identifies
the absence of a measurement start.
The zero timestamp is an initialized sentinel, not an uninitialized
C-language read. The failure is the absence of the measurement it should
represent.

\paragraph{H2: Premature consumption.}
A required predecessor update $U$ occurs too late:
\begin{equation}C\prec_h U\quad\text{instead of}\quad U\prec_h C.\end{equation}
The PELT case synchronizes the idle clock before the previous scheduling
class updates required signals.

\paragraph{H3: Stale consumption.}
An older value survives into an event requiring a refresh:
\begin{equation}S_{\mathrm{old}}\prec_h C\prec_h S_e.\end{equation}
The \utilest case observes CPU utilization before its update for the
current event. These hazards are useful, non-exhaustive review questions;
they need not be disjoint categories.

\subsection{Consumer Relocation}
Moving a site from $E_{C_0}$ to $E_{C_1}$ makes histories newly reachable:
\begin{equation}\Delta R=R(E_{C_1})\setminus R(E_{C_0}).\end{equation}
Admission additionally depends on $G$. Review must ask both whether the
newly reached events should consume the measurement and whether each
newly admitted consumption has compatible producer state.

\section{Linux Scheduler Background}
\subsection{Scheduling Domains and Search Cost}
Linux scheduling domains are platform- and configuration-dependent CPU
spans over which the scheduler searches for work or compares load.
SMT, MC, PKG, and NUMA are labels in the examined statistics, not a
universal fixed hierarchy or a one-to-one mapping to cache levels.
When a CPU exhausts local work, new-idle balancing can search remote
runqueues. The search can recover capacity but consumes scheduler time
and may lead to migration. We do not decompose unmeasured lock,
cache-coherence, or interconnect costs.

\subsection{Measurement Interval and State Lifecycle}\label{sec:interval}
We pin the source analysis to the correction revision and its original
accounting-change predecessor records \cite{idle_source_snapshot,idle_relocation_merge}.
The start of the measured interval is the assignment in
\newidle, before the balancing search. The source explicitly includes
that search duration in idle-time accounting. The interval is therefore
not identical to idle-task execution time, a hardware C-state residency,
or instantaneous CPU availability.

If balancing obtains runnable work, the end path clears the timestamp
without submitting an idle sample. If the scheduler proceeds into idle,
the established timestamp can survive until the idle-task switch-out
consumer. A path returning before the assignment can select the idle task
without starting this measurement. The helper's guard distinguishes these
histories rather than mechanically timing every idle-task dispatch.
Table~\ref{tab:lifecycle} records the relevant state transitions.

\begin{table*}[t]
\centering\small
\caption{Timestamp lifecycle in the pinned source revision
\cite{idle_source_snapshot}. Line ranges refer to the retained source snapshot.}
\label{tab:lifecycle}
\begin{tabularx}{\textwidth}{@{}YYY@{}}
\toprule\textbf{Event / source site}&\textbf{State transition}&\textbf{Measurement interpretation}\\\midrule
Runqueue initialization (core.c 9066--9068)&Timestamp starts at zero; estimator and cost have initial values.&Zero is a sentinel, not an uninitialized read.\\\addlinespace
Pending-wakeup bailout (fair.c 14555--14556)&Returns before timestamp establishment.&Idle-task selection may occur without a measured interval.\\\addlinespace
New-idle establishment (fair.c 14559--14563)&Timestamp receives the runqueue clock before search.&The interval includes balancing work if it survives into idle.\\\addlinespace
Work obtained (fair.c 14673--14674)&A nonzero pulled-task result clears the timestamp.&No continuing idle interval is submitted on this branch.\\\addlinespace
Idle-class switch-out (idle.c 483--488; core.c 3743--3758)&Guarded sample construction, estimator update, clamp, timestamp reset.&Consume a represented interval once; otherwise skip sampling.\\\bottomrule
\end{tabularx}
\end{table*}

\subsection{Idle History and Search Termination}\label{sec:policy}
Linux tracks historical idle duration and new-idle balancing cost.
Writing $A_k$ for \avgidle, the exponentially weighted update is
idealized without integer rounding as
\begin{equation}A_k=\frac{7}{8} A_{k-1}+\frac{1}{8}\Delta_{\mathrm{idle}}.\end{equation}
Let $C_d$ denote the relevant scheduling-domain new-idle cost and
$C_{\mathrm{spent}}$ the cost already incurred. Relevant comparisons
can be abstracted as
\begin{equation}\label{eq:firstgate}
\operatorname{TerminateSearch}\Leftarrow A<C_d,
\end{equation}
\begin{equation}\label{eq:secondgate}
\operatorname{AbortSearch}\Leftarrow A<C_{\mathrm{spent}}+C_d.
\end{equation}
These isolate \avgidle-dependent predicates, not the complete control
flow, which also depends on eligibility, pending wakeups, and other
conditions. Driving $A$ to its clamp can weaken these cost gates; passing
a gate does not guarantee a task will be found or migrated. In the pinned
source, the initial check also depends on root-domain overload (fair.c
14584); the per-domain check occurs at 14606. A separate NO\_HZ consumer
compares \avgidle with migration cost before requesting blocked-load
maintenance (14504). Other flags and randomized balancing can still
restrict search \cite{idle_source_snapshot}.

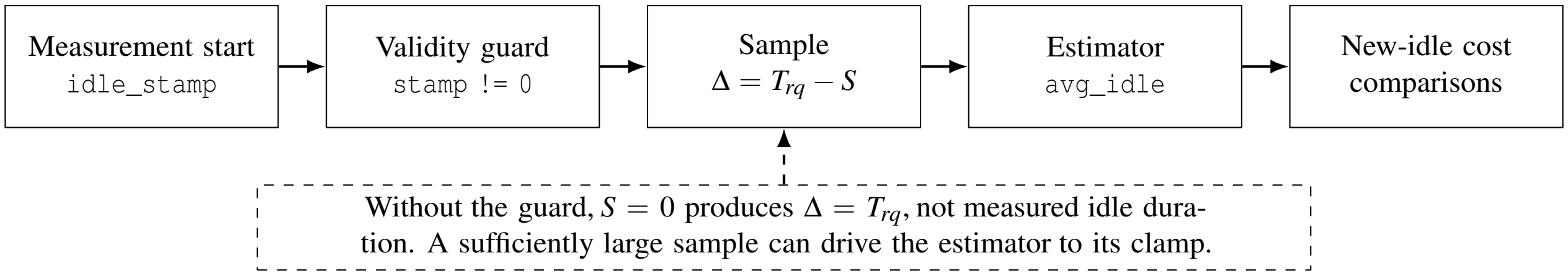
\begin{figure*}[t]
\centering
\begin{tikzpicture}[font=\small,
 box/.style={draw,align=center,text width=27mm,minimum height=13mm,inner sep=3pt},
 arr/.style={-{Latex[length=2mm]},thick},node distance=5mm]
\node[box] (s) {Measurement start\\\idlestamp};
\node[box,right=of s] (g) {Validity guard\\\texttt{stamp != 0}};
\node[box,right=of g] (d) {Sample\\$\Delta=T_{rq}-S$};
\node[box,right=of d] (a) {Estimator\\\avgidle};
\node[box,right=of a] (c) {New-idle cost\\comparisons};
\draw[arr] (s)--(g);\draw[arr] (g)--(d);\draw[arr] (d)--(a);\draw[arr] (a)--(c);
\node[draw,dashed,align=center,text width=110mm,below=6mm of d] (bad)
{Without the guard, $S=0$ produces $\Delta=T_{rq}$, not measured idle duration. A sufficiently large sample can drive the estimator to its clamp.};
\draw[arr,dashed] (bad.north)--(d.south);
\end{tikzpicture}
\caption{Measurement-to-policy path. The final guard acts before sample
construction, not at the downstream cost comparisons. Search remains
subject to other scheduler conditions.}\label{fig:control_loop}
\end{figure*}

\section{Closing the Coverage Gap: Helper Introduction and Relocation}\label{sec:coverage}
\subsection{The Original Wakeup-Side Consumer}
Before relocation, the relevant accounting was attached to a wakeup path:
\begin{lstlisting}[language=C,caption={Simplified original accounting.}]
if (rq->idle_stamp) {
    u64 delta = rq_clock(rq) - rq->idle_stamp;
    update_avg(&rq->avg_idle, delta);
    if (rq->avg_idle > 2 * rq->max_idle_balance_cost)
        rq->avg_idle = 2 * rq->max_idle_balance_cost;
    rq->idle_stamp = 0;
}
\end{lstlisting}
It both restricted the arrival paths reaching consumption and guarded
sample construction with a nonzero timestamp.

\subsection{The Coverage Defect}
Task creation, exec-related placement, and load-balancing migration were
identified as arrival classes that could bypass the wakeup-side site
\cite{idle_coverage}. Their exact control flow differs; the shared
property is an idle exit without executing that accounting point.
This does not mean every fork, exec, or migration has this behavior.
Omitted samples make the accounting incomplete, but the direction of
estimator bias depends on the intervals and existing history.
Table~\ref{tab:coverage} identifies concrete source chains. These are
source-level arrival mechanisms, not newly collected runtime traces.
The relevant case is a destination already executing the idle task with
an established measurement, followed by arrival and a switch-out; merely
executing fork, exec, or migration is not sufficient.

\begin{table*}[t]
\centering\small
\caption{Arrival-path coverage audit in the pinned source
\cite{idle_source_snapshot}. The old accounting site is
\texttt{ttwu\_do\_activate()}; the relocated site is idle-class switch-out.}
\label{tab:coverage}
\begin{tabularx}{\textwidth}{@{}>{\RaggedRight\arraybackslash}p{0.16\textwidth}YY@{}}
\toprule\textbf{Arrival class}&\textbf{Destination activation chain}&\textbf{Coverage condition and limit}\\\midrule
Ordinary wakeup&\texttt{ttwu\_do\_activate()} $\rightarrow$ \texttt{activate\_task()}.&Reaches the original wakeup-side site; relocated accounting instead consumes at idle-class switch-out.\\\addlinespace
New task&\texttt{wake\_up\_new\_task()} $\rightarrow$ \texttt{activate\_task()} (core.c 4957--4979).&Direct activation bypasses the old site. A later idle-class switch-out reaches the helper if this arrival ends an established idle interval.\\\addlinespace
Load-balance attachment&\texttt{attach\_tasks()} $\rightarrow$ \texttt{attach\_task()} $\rightarrow$ \texttt{activate\_task()} (fair.c 11164--11177; sched.h 3122--3128).&Direct destination attachment bypasses the wakeup-side site. Successful local new-idle pulling can clear the stamp instead of entering idle.\\\addlinespace
Exec-related migration&\texttt{sched\_exec()} selects another CPU; stopper migration can proceed through \texttt{\_\_migrate\_task()} $\rightarrow$ \texttt{move\_queued\_task()} $\rightarrow$ \texttt{activate\_task()}.&Conditional migration path, not every exec. Destination activation bypasses the old site; affinity, task state, and intervening scheduler activity constrain whether it ends a measured idle interval.\\\bottomrule
\end{tabularx}
\end{table*}

\subsection{Extracting the Helper and Moving the Boundary}
The change introduced \updaterq and invoked it at \putprevidle:
\begin{lstlisting}[language=C,caption={Simplified switch-out call site; the signature is abbreviated.},label={lst:switchout}]
static void put_prev_task_idle(struct rq *rq, ...) {
    update_curr_idle(rq);
    scx_update_idle(rq, false, true);
    update_rq_avg_idle(rq);
}
\end{lstlisting}
Intermediate revisions explored \texttt{activate\_task()} and
\texttt{enqueue\_task()}; v7 moved the helper to idle-task switch-out,
and v8 removed the timestamp guard \cite{idle_relocation}. The design
thus changed the organizing event from a particular task-arrival
mechanism to a common idle-exit boundary. Activation-site alternatives
track arrival rather than actual idle-class switch-out and must account
for arrivals that do not produce that transition. Conversely, switch-out
centralizes completion but expands the histories reaching the consumer,
requiring an explicit measurement-validity check. These are design
trade-offs exposed by the source structure, not measurements comparing
all intermediate patch revisions. The original patch-submission
measurements reported benefits on the 80-core Ampere system, and the
change was integrated into the scheduler tree
\cite{idle_relocation,idle_relocation_merge}.

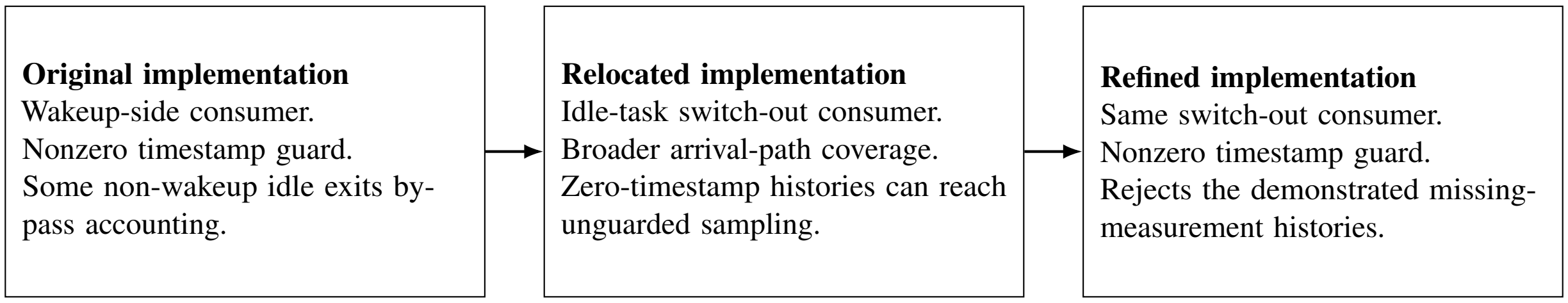
\begin{figure*}[t]
\centering
\begin{tikzpicture}[font=\small,
 box/.style={draw,align=left,text width=46mm,minimum height=30mm,inner sep=5pt},
 node distance=6mm,arr/.style={-{Latex[length=2mm]},thick}]
\node[box] (old) {\textbf{Original implementation}\\Wakeup-side consumer.\\Nonzero timestamp guard.\\Some non-wakeup idle exits bypass accounting.};
\node[box,right=of old] (move) {\textbf{Relocated implementation}\\Idle-task switch-out consumer.\\Broader arrival-path coverage.\\Zero-timestamp histories can reach unguarded sampling.};
\node[box,right=of move] (fix) {\textbf{Refined implementation}\\Same switch-out consumer.\\Nonzero timestamp guard.\\Rejects the demonstrated missing-measurement histories.};
\draw[arr] (old)--(move);\draw[arr] (move)--(fix);
\end{tikzpicture}
\caption{Evolution of the accounting boundary across three implementation states.
The relocation addresses coverage; the refinement addresses validity
without restoring wakeup-only accounting.}\label{fig:reachability_sets}
\end{figure*}

\section{Behavioral Evidence from Upstream Testing}\label{sec:evidence}
\subsection{Independent Scheduler Statistics}
K.~Prateek Nayak's review reported no broad regression, but recurring
tbench-256 sensitivity. The initial mean-throughput comparison was
$-5\%$; profiling reruns were closer to approximately $-2\%$, described
as within the margin of error. The reviewer noted tbench's short sleeps
and client/server locality within an LLC domain
\cite{idle_relocation_review}.

The profiling windows lasted 20,402 and 20,396 jiffies. Table~\ref{tab:schedstats}
reports counts over those nearly equal windows, not universal rates.
\begin{table}[t]
\centering\small
\caption{Independent review statistics. Success counts are derived by the
profiling tool \cite{idle_relocation_review}.}\label{tab:schedstats}
\begin{tabular}{@{}lrrrr@{}}\toprule
&\multicolumn{2}{c}{Attempts}&\multicolumn{2}{c}{Derived successes}\\
\cmidrule(lr){2-3}\cmidrule(l){4-5}
Domain&Before&After&Before&After\\\midrule
SMT&3,964&17,961&189&1,011\\
MC&1,917&7,022&56&218\\
PKG&43&13&1&0\\
NUMA&21&4&1&1\\\bottomrule
\end{tabular}
\end{table}
SMT attempts increased approximately $4.53\times$ and MC attempts
$3.66\times$. Derived success fractions rose from 4.77\% to 5.63\% at
SMT and from 2.92\% to 3.10\% at MC. PKG and NUMA attempts decreased.
The derived success counts are not direct task-migration counts: the
baseline PKG record reports one derived success but zero tasks gained.
Their derivation matters when interpreting balancing outcomes.

The activity change is therefore topology dependent. It does not establish
that a particular invalid timestamp update caused the application-level
throughput difference, nor that additional search is always useful.

\subsection{Measurement-to-Policy Path}
For compatible establishment,
\begin{equation}\Delta_{\mathrm{valid}}=T_{rq}-S_{\mathrm{idle}}.\end{equation}
The initial relocated helper instead performed unconditional subtraction:
\begin{lstlisting}[language=C,caption={Simplified initial helper.},label={lst:unguarded}]
void update_rq_avg_idle(struct rq *rq) {
    u64 delta = rq_clock(rq) - rq->idle_stamp;
    u64 max = 2 * rq->max_idle_balance_cost;
    update_avg(&rq->avg_idle, delta);
    if (rq->avg_idle > max)
        rq->avg_idle = max;
    rq->idle_stamp = 0;
}
\end{lstlisting}
With a zero timestamp,
\begin{equation}\Delta_{\mathrm{bad}}=T_{rq}.\end{equation}
Zhan Xusheng's review explains that roughly one eighth of the input
difference is incorporated into the estimate. A sufficiently large
runqueue clock can therefore drive \avgidle to its much smaller clamp
on an incompatible update \cite{idle_saturation_review}.
The problem is estimator saturation, not necessarily integer overflow.
Ignoring integer rounding, with old estimate $A$, sample $T$, and clamp
$M=2\maxidlecost$, saturation occurs when $7A+T\geq8M$.
The actual update uses a signed difference and integer division, so this
expression is an explanatory condition within the ordinary representable
range, not a universal arithmetic proof.
The source-level control dependency in Figure~\ref{fig:control_loop}
and independent workload statistics are distinct kinds of evidence.

\section{Remaining Controlled Validation}\label{sec:protocol}
The historical record establishes correctness reachability and behavioral
sensitivity, but does not isolate the guard's performance effect across
three accounting states on one common kernel baseline.
A controlled evaluation should compare:
\begin{enumerate}[leftmargin=*,nosep]
\item \textbf{Wakeup-side baseline:} guarded accounting with the original
coverage limitation;
\item \textbf{Relocated-only:} broader switch-out accounting without the
validity guard;
\item \textbf{Final guarded:} the same relocation with nonzero timestamp
validation.
\end{enumerate}
Use the same non-accounting kernel source, compiler, configuration,
firmware, frequency policy, and benchmark harness. Record the exact
accounting-only changes; comparing January and September snapshots
would confound the result with unrelated scheduler development.

Prioritize the original hackbench and schbench configurations, tbench
around the sensitive concurrency range, and a compute-bound control.
Document exact commands and what threads, groups, and clients mean.
Repeat runs sufficiently to report variability and justified confidence
intervals rather than a best result. Measure runtime or throughput as
defined by each harness, tail latency where applicable, migrations, and
per-domain attempts and balancing outcomes.

Instrumentation should record attempted updates, timestamp validity,
sample values, estimator before and after, and the contemporaneous clamp.
Directed arrival-path tests should establish whether accounting is
missed; zero-stamp traces should distinguish invalid consumption from
valid, short idle intervals. Identifying the \ttwupending cause requires
path context, not only a zero-timestamp count. Instrumentation availability
and overhead must be checked rather than assuming a particular tracepoint.

The questions are whether relocation closes the identified missing-update
paths and how much behavioral change comes from coverage versus invalid
sampling. This protocol specifies evidence still needed; it is not a
completed experiment or a new performance result.

\section{Missing Establishment After Relocation}\label{sec:validity}
\subsection{Independent Diagnosis and Runtime Reachability}
Masahito S. independently proposed restoring the guard and adding an
idle-entry timestamp in \setnextidle
\cite{idle_symmetric_proposal}. The eventual correction retains the
first component but not the symmetric-entry change.

Temporary tracing reported in the correction thread confirmed entry to \updaterq with
\texttt{idle\_stamp == 0}. The correction thread reports no material
hackbench regression against its stated comparison point
\cite{idle_guard_v3}. The follow-up report describes repeated thread/process cases with mean
deltas within approximately 1.5\% and similar baseline variability
\cite{idle_guard_followup}. Neither a trace file nor per-run measurements
is supplied with this manuscript. The evidence is therefore an attributed
observation, not an independently reproduced trace or a statistical
equivalence result.

\subsection{The Pending-Wakeup History}
\begin{figure}[t]
\centering
\begin{tikzpicture}[font=\small,node distance=4mm,
 box/.style={draw,align=left,text width=68mm,inner sep=5pt},
 arr/.style={-{Latex[length=2mm]},thick}]
\node[box] (n) {\textbf{1. New-idle entry}\\Pending wakeup is already recorded.};
\node[box,below=of n] (b) {\textbf{2. Early return}\\\texttt{ttwu\_pending} bailout precedes the ordinary timestamp assignment.};
\node[box,below=of b] (i) {\textbf{3. Idle-task execution}\\The runqueue can enter idle with a zero timestamp.};
\node[box,below=of i] (w) {\textbf{4. Pending wakeup processed}\\Switch-out reaches the relocated helper.};
\node[box,below=of w] (g) {\textbf{5. Measurement boundary}\\Unguarded: sample is $T_{rq}$.\\Refined: zero timestamp skips sampling.};
\draw[arr] (n)--(b);\draw[arr] (b)--(i);\draw[arr] (i)--(w);\draw[arr] (w)--(g);
\end{tikzpicture}
\caption{A reachable pending-wakeup history, not a deterministic workload
trace. The validity guard separates reaching the consumer from constructing
a measured idle-duration sample.}
\label{fig:pending-flow}
\end{figure}
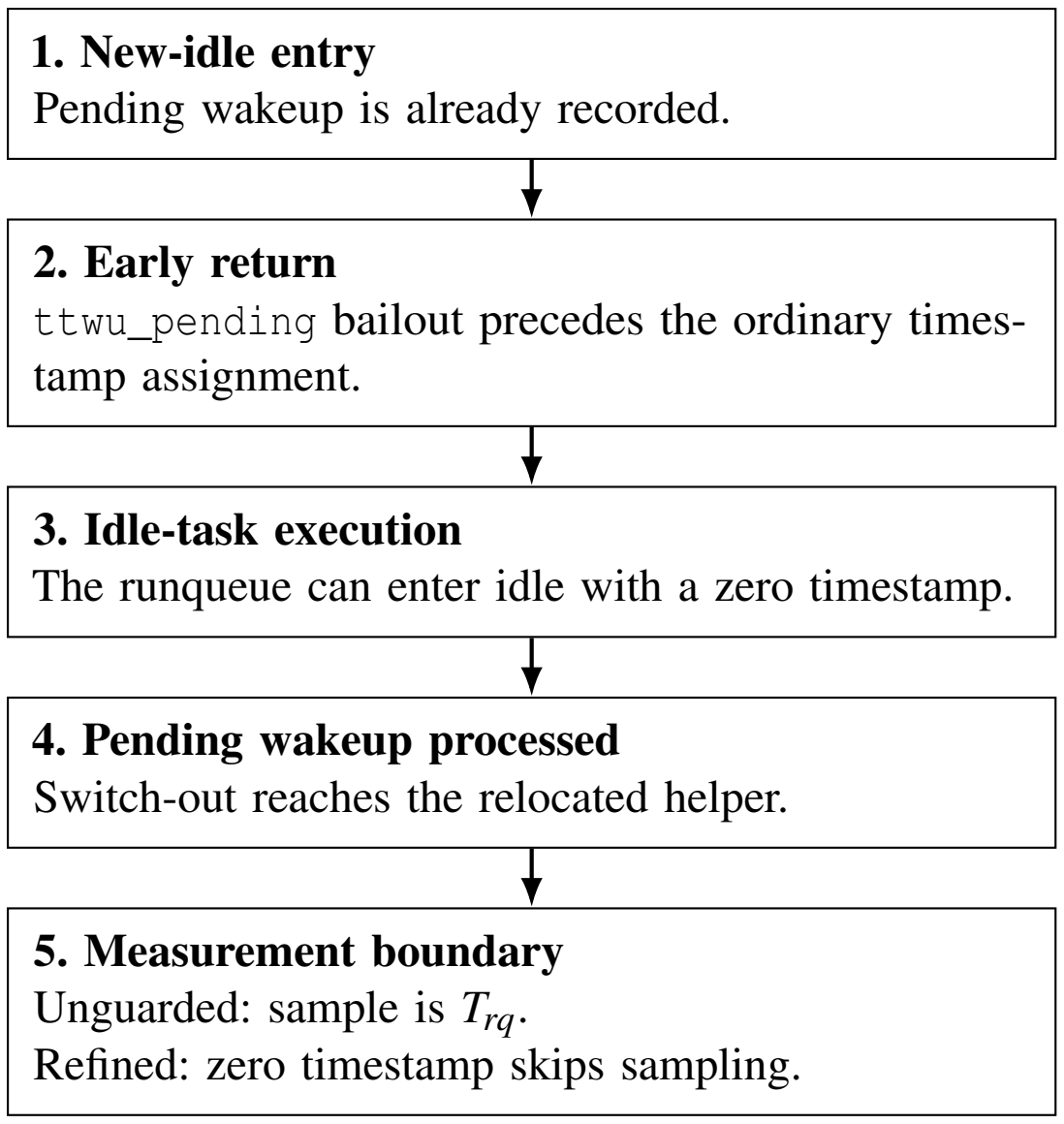

Review identified the relevant ordering in \newidle
\cite{idle_source_snapshot}:
\begin{lstlisting}[language=C,caption={Simplified ordering of bailout and timestamp establishment.},label={lst:ttwu}]
if (this_rq->ttwu_pending)
    return 0;
/* ... */
this_rq->idle_stamp = rq_clock(this_rq);
\end{lstlisting}
A runqueue can return before the assignment, execute the idle task,
process the pending wakeup, and switch away from idle. The relocated
consumer then executes without the expected measurement start.
Figure~\ref{fig:pending-flow} summarizes this ordering. For such a history,
\begin{equation}h\in R(E_{C_1}),\qquad h\notin R_S(E_{C_1}).\end{equation}
Unconditional admission makes $h\in R_{G_1}(C_1)$, violating
Eq.~\ref{eq:safety}. Zero is not an exceptionally short measurement;
the compatible establishment did not occur.

\subsection{A Class of Histories, Not One Exception}
The ordinary establishment point is below the bailout. The relevant
class includes histories reaching the consumer while missing that point
\cite{idle_saturation_review}. Review must ask which newly reachable
histories possess measurement state, rather than only which individual
function appears to omit initialization.

\subsection{Why Symmetric Entry Is Insufficient}
Timestamping every idle-task entry would appear to pair entry with
switch-out. But the correction identifies \texttt{find\_proxy\_task()}
and force-idling as cases where idle-task selection need not establish
the ordinary measurement interval
\cite{idle_guard_v3,idle_symmetric_review}. Thus
\begin{equation}
\text{idle task selected}\centernot{\Rightarrow}\text{compatible }S.
\end{equation}
This is a semantic distinction, not a new claim about security, physical
core residency, cache behavior, or energy consumption.

\section{Preserving Coverage While Refining Validity}\label{sec:repair}
The final correction retains the broader consumer and filters sampling:
\begin{lstlisting}[language=C,caption={Essential structure of the correction.},label={lst:fix}]
void update_rq_avg_idle(struct rq *rq) {
    u64 idle_stamp = rq->idle_stamp;
    u64 delta, max;
    if (!idle_stamp)
        return;
    delta = rq_clock(rq) - idle_stamp;
    update_avg(&rq->avg_idle, delta);
    max = 2 * rq->max_idle_balance_cost;
    if (rq->avg_idle > max)
        rq->avg_idle = max;
    rq->idle_stamp = 0;
}
\end{lstlisting}
The concrete admission predicate is
\begin{equation}G_{\mathrm{impl}}(h):\mathtt{idle\_stamp}(h)\ne0.\end{equation}
It excludes the observed zero-timestamp histories. It is not a general
proof that every nonzero timestamp is compatible under all scheduler
changes. In the pinned core/fair source, the ordinary nonzero assignment
is in new-idle balancing; initialization, successful pulling, and helper
completion reset the field. A history that misses the assignment after
one of those resets therefore reaches sampling with zero. This supports
the demonstrated failure class while leaving future changes subject to
provenance and ordering review. The site stays broad; admitted sampling
histories become narrower.

The timestamp also has placement readers: fair.c 8482--8491 uses recent
idle entry as a tie-breaker among CPUs already checked as available.
Changing its reset point can change the lifetime seen by those readers.
We identify this additional audit obligation but do not claim a measured
placement effect. A stale comment mentioning \avgidle in
\texttt{select\_idle\_cpu()} is not treated as an active dependency: the
inspected body instead uses the SIS\_UTIL scan budget
\cite{idle_source_snapshot}.

The correction commit records reviews from K.~Prateek Nayak
and Vincent Guittot and an acknowledgment from John Stultz. Zhan Xusheng
additionally supplied a Reviewed-by in the discussion; that tag is not
listed in the merged-commit trailers
\cite{idle_guard_v3,vincent_guard_review,idle_saturation_review,idle_guard_merge}.
Both changes are available in the public Linux mainline repository.
The commit records establish implementation and authorship, but this paper
does not infer first-containing release tags or stable backports from
commit timestamps \cite{idle_relocation_merge,idle_guard_merge}.

Figure~\ref{fig:reachability_sets} distinguishes the three actual upstream
implementation states: guarded but incomplete, broader but admitting
missing-establishment samples, and broader with validity filtering.
The final guard does not undo the original coverage intervention.

\section{Cross-Case Analysis of Scheduler State Lifetimes}\label{sec:independent}
Two independent cases test whether the lifetime questions coherently
describe other scheduler failures. This retrospective analysis demonstrates
applicability, not predictive accuracy or automated bug finding.

\begin{table*}[t]\centering\small
\caption{Retrospective mapping to the lifetime audit.}\label{tab:crosscases}
\begin{tabularx}{\textwidth}{@{}>{\RaggedRight\arraybackslash}p{0.13\textwidth}YYY@{}}
\toprule Case&Required state&Violation&Repair\\\midrule
\avgidle&Timestamp for the intended idle interval.&Sampling occurs without measurement establishment (H1).&Keep switch-out accounting; reject zero timestamps.\\\addlinespace
PELT&Previous-class PELT signals updated before idle synchronization.&Synchronization precedes the required update (H2).&Move synchronization after previous-class processing.\\\addlinespace
\utilest/\schedutil&CPU estimate current for enqueue or dequeue.&Frequency-policy observation sees an older estimate (H3).&Update CPU estimate before notification; retain task update at its dependency-correct point.\\\bottomrule
\end{tabularx}
\end{table*}

\subsection{Premature PELT Synchronization}
The PELT correction states that fair scheduling synchronized its idle
clock when \picknextfair failed to select a task, before the previous
scheduling class could update required signals. RT-runqueue
\texttt{util\_avg} was affected \cite{pelt_idle_fix}.
Define $S_{\mathrm{PELT}}$ as the previous-class update and
$C_{\mathrm{PELT}}$ as \peltidleclock. The old ordering was
$C_{\mathrm{PELT}}\prec S_{\mathrm{PELT}}$; the repair moved
synchronization to \setnextidle after previous-class processing.
The source credits Samuel Wu and Alex Hoh with reporting/testing,
Vincent Guittot with authorship, and Peter Zijlstra with committing
\cite{pelt_idle_fix}. This is H2.

\subsection{Stale CPU Utilization at Schedutil}
\utilest supplements PELT for utilization tracking
\cite{util_est_intro}. During enqueue, \updateloadavg could indirectly
notify \schedutil before \utilestenqueue made CPU utilization current.
The correction moved that estimate update earlier in \enqueuetaskfair
\cite{util_est_enqueue_fix}.

A complementary dequeue correction addressed CPU utilization being used
for frequency selection before its update \cite{util_est_dequeue_fix}.
The ordering was
\begin{equation}
S_{\mathrm{old}}\prec C_{\mathrm{DVFS}}\prec S_{\mathrm{CPU,current}}.
\end{equation}
The repair separated CPU and task updates: CPU state moves before the
notification path, whereas task state remains later because it depends
on updated task \texttt{util\_avg}. Initialized state was nevertheless
stale for the triggering event. This is H3, not a new measurement of
frequency or energy effects.

\section{Operational Lessons}
\subsection{Semantic Lifetime Is Not Source Proximity}
The main case distinguishes task-state transitions from measurement
boundaries. PELT distinguishes entering idle from completing the previous
class's updates. \utilest distinguishes two estimates manipulated
together but subject to different dependencies. Locally plausible
source placement is not sufficient to establish compatibility.

\subsection{Search Activity Is Not Search Utility}
The independent results show several-fold increases in SMT and MC
attempts with more modest changes in success fractions, and decreases
at PKG and NUMA. Additional balancing is not automatically beneficial.
The data show sensitivity and topology dependence, not a causal
allocation of throughput effects to individual invalid samples.

\subsection{A Review Checklist}
For each stateful policy consumer:
\begin{enumerate}[leftmargin=*,nosep]
\item Identify the event, interval, or history the state should represent.
\item Enumerate histories reaching the consumer, including those added by relocation.
\item Locate compatible establishment or update for each admitted history.
\item Check missing establishment, premature consumption, and stale reuse.
\item Determine what the implementation guard excludes and what semantic
obligations remain.
\end{enumerate}
This checklist complements testing, race analysis, and verification; it
does not replace them. The \avgidle path affects topology search, while
the \utilest path reaches frequency policy. Neither control dependency
alone establishes an application-level, energy, or hardware benefit.

\begin{table*}[t]
\centering\small
\caption{Evidence supports distinct conclusions. Historical observations are
not substituted for the unperformed controlled comparison.}
\label{tab:evidence-map}
\begin{tabularx}{\textwidth}{@{}YYY@{}}
\toprule
\textbf{Evidence} & \textbf{Supported conclusion} & \textbf{Boundary}\\
\midrule
Original Ampere patch tests & Configuration-specific hackbench improvement and schbench tail-latency observations. & No general benefit or isolated effect of the later guard.\\\addlinespace
Independent AMD review & Topology-dependent balancing activity and tbench sensitivity. & No per-sample causal attribution to application throughput.\\\addlinespace
Temporary tracing and source ordering & Zero-timestamp consumption is reachable; the bailout precedes establishment. & A zero count alone does not identify the triggering path.\\\addlinespace
Merged helper refinement & The nonzero guard rejects demonstrated missing-establishment samples. & Nonzero alone is not a universal proof of semantic compatibility.\\\addlinespace
Proposed three-variant experiment & Defines how coverage and invalid-sample effects could be separated. & Not executed or reported as results in this manuscript.\\
\bottomrule
\end{tabularx}
\end{table*}

\section{Evidence Boundaries}
Table~\ref{tab:evidence-map} distinguishes the conclusions supported by
each evidence source and the remaining attribution limits.

\paragraph{Performance attribution.}
The historical tests are not a controlled three-variant experiment.
Patch-submission hackbench/schbench observations and independent AMD
results have distinct provenance. Neither is recast as a new experiment. The 7.2\% result lacks raw values,
commands, trial counts, and an explicit percentage definition in the cited
commit record; this paper reports it as stated rather than converting it
into a new throughput measurement. The AMD profile is one reported
comparison and does not establish a distribution across machines or runs.
No individual sample is mapped to a particular migration, cache event,
or application latency.

\paragraph{Guard sufficiency.}
The zero guard excludes demonstrated missing-establishment histories.
Full compatibility additionally requires reasoning about nonzero
state's provenance, ordering, and freshness.

\paragraph{Cross-case scope.}
PELT and \utilest are retrospective source analyses, not new
reproductions. The taxonomy is non-exhaustive and has no measured
precision or recall. Synchronization, races, atomicity, representation,
arithmetic, and multi-variable consistency can introduce other failures.
$\operatorname{Compat}$ is a manual review obligation, not an automated
implementation.

\section{Related Work}
\paragraph{Multicore scheduling and placement.}
Lozi et~al. analyze Linux balancing and placement inefficiencies
\cite{lozi2016wasted}. Arachne emphasizes core-aware thread management
\cite{arachne2018}; Shenango and Shinjuku explore low-latency scheduling
and allocation for server workloads \cite{shenango2019,shinjuku2019}.
The present case concerns the semantic validity of historical state
consumed by kernel balancing policy.

\paragraph{Feedback and estimates.}
Feedback-driven CPU allocation and dependency-sensitive scheduling
illustrate measurements guiding resource policy
\cite{steere1999feedback,zheng2004swap}. Linux \utilest adds an estimate
on top of PELT \cite{util_est_intro}; its corrections show why updates
must precede downstream policy observation
\cite{util_est_enqueue_fix,util_est_dequeue_fix}.

\paragraph{Verification and instrumentation.}
Coccinelle documents and automates collateral kernel changes
\cite{coccinelle2008}; longitudinal fault studies and system-specific
checking provide broader bug-analysis methods
\cite{palix2011faults,engler2000checker}. seL4 and CertiKOS establish
stronger verified-kernel guarantees \cite{klein2009sel4,gu2016certikos}.
DTrace and KernInst provide kernel instrumentation
\cite{cantrill2004dtrace,tamches1999kerninst}, while Aragog addresses
runtime verification of networked systems \cite{yaseen2020aragog}.
Our contribution is narrower: an operational intervention and a semantic
audit abstraction for stateful policy refactorings, not a new checker.

\section{Conclusion}
\label{sec:conclusion}

Modern operating system schedulers increasingly depend on historical heuristics 
to navigate complex multicore and NUMA topologies. When such signals govern 
software control paths over hardware resources, their measurement validity 
becomes integral to policy correctness. 

This paper presented the operational lifecycle of a targeted architectural 
redesign in the Linux scheduler. By introducing \updaterq and relocating 
accounting from wakeup-specific paths to \putprevidle, the upstream kernel 
closed a longstanding coverage defect that overlooked task arrivals from 
fork/clone, \texttt{execve}, and load-balancing migrations. Mainline deployment, 
however, exposed a subtle state-lifetime hazard: decoupling consumption from 
wakeups permitted \ttwupending bails to reach switch-out with an uninitialized 
timestamp, injecting raw uptime into the estimator and saturating \avgidle. 
The subsequent mainline correction (commit \texttt{c6dcd97c8be75}) hardened 
this interface by restoring an explicit validity guard, preserving universal 
idle-exit coverage while rejecting invalid samples.

The fundamental operational lesson is that\ \emph{control-flow symmetry does not 
imply measurement-lifetime symmetry}. Mechanically pairing idle-task entry and 
exit hooks fails in modern operating systems because mechanisms such as SMT Core 
Scheduling force-idling and proxy execution dispatch the idle thread for 
isolation or locking reasons without establishing a genuine physical idle 
interval. Stateful policy inputs must therefore be audited not merely for 
syntactic reachability, but for semantic epoch compatibility. As demonstrated 
by our retrospective analysis of PELT idle-clock synchronization and 
\utilest/\schedutil frequency selection, the hazards of missing establishment 
($H_1$), premature observation ($H_2$), and stale consumption ($H_3$) represent 
recurrent failure modes across scheduler and DVFS boundaries.

While isolating the granular microarchitectural performance consequences across 
historical kernel revisions remains an important area for controlled three-variant 
experimentation, the hardened switch-out architecture provides an immediate, 
robust foundation for production Linux load balancing. Looking forward, we believe 
the state-lifetime criteria formalized in this study can be operationalized into 
automated static analysis passes (e.g., via Coccinelle or compiler-level typestate 
checkers), helping systems programmers verify that kernel heuristics observe 
semantically valid state before guiding critical placement and power decisions.

\begingroup
\RaggedRight
\setlength{\emergencystretch}{3em}
\interlinepenalty=1000
\bibliographystyle{plain}
\bibliography{refs}

@misc{idle_coverage,
 author={Kaushik, Shubhang},
 title={{Idle-time accounting series: revision history through v8}},
 year={2026}, month=jan,
 howpublished={Linux Kernel Mailing List. \href{https://lore.kernel.org/all/20260121-v8-patch-series-v8-1-b7f1cbee5055@os.amperecomputing.com/}{Series and revision history}},
 note={January 21; includes links to the 2025 v1--v7 submissions}
}

@misc{idle_relocation,
 author={Kaushik, Shubhang},
 title={{[PATCH v8] sched: update rq$\rightarrow$avg\_idle when a task is moved to an idle CPU}},
 year={2026}, month=jan,
 howpublished={Linux Kernel Mailing List. \href{https://lore.kernel.org/all/20260121-v8-patch-series-v8-1-b7f1cbee5055@os.amperecomputing.com/}{Source record}},
 note={January 21; original helper introduction and accounting relocation}
}

@misc{idle_relocation_review,
 author={Nayak, K. Prateek},
 title={{Re: [PATCH v8] sched: update rq$\rightarrow$avg\_idle when a task is moved to an idle CPU}},
 year={2026}, month=jan,
 howpublished={Linux Kernel Mailing List. \href{https://lore.kernel.org/all/953b0294-312e-4f58-a459-43623a0f5128@amd.com/}{Source record}},
 note={January 22; independent tbench results and attached scheduler statistics}
}

@misc{idle_relocation_merge,
 author={Kaushik, Shubhang},
 title={{sched: Update rq$\rightarrow$avg\_idle when a task is moved to an idle CPU}},
 year={2026}, month=jan,
 howpublished={Linux mainline commit. \href{https://kernel.googlesource.com/pub/scm/linux/kernel/git/torvalds/linux/+/4b603f1551a73e2868b9e7a14b3938c23275cefb}{Commit record}},
 note={Commit 4b603f1551a73e2868b9e7a14b3938c23275cefb; committed January 22 by Peter Zijlstra}
}

@misc{idle_symmetric_proposal,
 author={{Masahito S.}},
 title={{[PATCH v2] sched/idle: Fix avg\_idle saturation by establishing symmetric idle entry hook}},
 year={2026}, month=apr,
 howpublished={Linux Kernel Mailing List. \href{https://lists.openwall.net/linux-kernel/2026/04/23/50}{Source record}},
 note={Independent proposal and saturation analysis}
}

@misc{idle_symmetric_review,
 author={Nayak, K. Prateek},
 title={{Re: [PATCH v2] sched/idle: Fix avg\_idle saturation by establishing symmetric idle entry hook}},
 year={2026}, month=apr,
 howpublished={Linux Kernel Mailing List. \href{https://lore.kernel.org/all/ca540897-7ca6-4540-b6f9-73bff98a2e54@amd.com/}{Source record}},
 note={April 23; review of symmetric-entry accounting}
}

@misc{idle_guard_v3,
 author={Kaushik, Shubhang},
 title={{[PATCH v3] sched/core: Skip rq$\rightarrow$avg\_idle update without a valid idle\_stamp}},
 year={2026}, month=aug,
 howpublished={Linux Kernel Mailing List. \href{https://lore.kernel.org/all/20260807-master-v3-1-c328354efed3@gentwo.org/}{Source record}},
 note={August 7; includes reported temporary tracing and workload comparison}
}

@misc{idle_saturation_review,
 author={Zhan, Xusheng},
 title={{Re: [PATCH v3] sched/core: Skip rq$\rightarrow$avg\_idle update without a valid idle\_stamp}},
 year={2026}, month=sep,
 howpublished={Linux Kernel Mailing List. \href{https://lore.kernel.org/all/20260901024438.1830934-1-zhanxusheng@xiaomi.com/}{Source record}},
 note={September 1; timestamp lifecycle and saturation analysis}
}

@misc{vincent_guard_review,
 author={Guittot, Vincent},
 title={{Re: [PATCH v3] sched/core: Skip rq$\rightarrow$avg\_idle update without a valid idle\_stamp}},
 year={2026}, month=aug,
 howpublished={Linux Kernel Mailing List. \href{https://lore.kernel.org/all/CAKfTPtBywWojCTedRR17_cLjbD0kkHWwjs9=dWL9Zh792-dPxQ@mail.gmail.com/}{Source record}},
 note={August 18; Reviewed-by}
}

@misc{idle_guard_merge,
 author={Kaushik, Shubhang},
 title={{sched/core: Skip rq$\rightarrow$avg\_idle update without a valid idle\_stamp}},
 year={2026}, month=sep,
 howpublished={Linux mainline commit. \href{https://kernel.googlesource.com/pub/scm/linux/kernel/git/torvalds/linux/+/c6dcd97c8be75f052a1ca52cf79b03e7292962f1}{Commit record}},
 note={Commit c6dcd97c8be75f052a1ca52cf79b03e7292962f1; committed September 2 by Peter Zijlstra}
}

@misc{pelt_idle_fix,
 author={Guittot, Vincent},
 title={{sched/fair: Fix pelt clock sync when entering idle}},
 year={2026}, month=jan,
 howpublished={Linux kernel commit. \href{https://kernel.googlesource.com/pub/scm/linux/kernel/git/sj/linux/+/98c88dc8a1ace642d9021b103b28cba7b51e3abc}{Source record}},
 note={Commit 98c88dc8a1ace642d9021b103b28cba7b51e3abc; fixes 17e3e88ed0b6}
}

@misc{util_est_intro,
 author={Bellasi, Patrick},
 title={{sched/fair: Add util\_est on top of PELT}},
 year={2018}, month=mar,
 howpublished={Linux kernel development. \href{https://lkml.iu.edu/1803.2/04702.html}{Source record}}
}

@misc{util_est_enqueue_fix,
 author={Bellasi, Patrick},
 title={{sched/fair: Update util\_est before updating schedutil}},
 year={2018}, month=may,
 howpublished={Linux kernel commit. \href{https://lkml.indiana.edu/hypermail/linux/kernel/1805.3/01693.html}{Source record}},
 note={Commit 2539fc82aa9b07d968cf9ba1ffeec3e0416ac721}
}

@misc{util_est_dequeue_fix,
 author={Yan, Xuewen},
 title={{sched/fair: Avoid stale CPU util\_est value for schedutil in task dequeue}},
 year={2021}, month=jan,
 howpublished={Linux kernel commit. \href{https://lkml.iu.edu/2101.1/07417.html}{Source record}},
 note={Commit 8c1f560c1ea3f19e22ba356f62680d9d449c9ec2}
}

@inproceedings{lozi2016wasted,
 author={Lozi, Jean-Pierre and Lepers, Baptiste and Funston, Justin and Gaud, Fabien and Qu{\'e}ma, Vivien and Fedorova, Alexandra},
 title={{The Linux Scheduler: A Decade of Wasted Cores}},
 booktitle={Proceedings of the Eleventh European Conference on Computer Systems (EuroSys '16)},
 year={2016}, pages={1--16}, publisher={Association for Computing Machinery},
 doi={10.1145/2901318.2901326},
 note={\href{https://doi.org/10.1145/2901318.2901326}{Publisher version}}
}

@inproceedings{arachne2018,
 author={Qin, Henry and Li, Qian and Speiser, Jacqueline and Kraft, Peter and Ousterhout, John},
 title={{Arachne: Core-Aware Thread Management}},
 booktitle={13th USENIX Symposium on Operating Systems Design and Implementation (OSDI 18)},
 year={2018}, pages={145--160}, publisher={USENIX Association},
 note={\href{https://www.usenix.org/conference/osdi18/presentation/qin}{Publisher version}}
}

@inproceedings{shenango2019,
 author={Ousterhout, Amy and Fried, Joshua and Behrens, Jonathan and Belay, Adam and Balakrishnan, Hari},
 title={{Shenango: Achieving High CPU Utilization for Latency-Sensitive Datacenter Workloads}},
 booktitle={16th USENIX Symposium on Networked Systems Design and Implementation (NSDI 19)},
 year={2019}, pages={505--520}, publisher={USENIX Association},
 note={\href{https://www.usenix.org/conference/nsdi19/presentation/ousterhout}{Publisher version}}
}

@inproceedings{shinjuku2019,
 author={Kaffes, Kostis and Chong, Timothy and Humphries, Jack Tigar and Belay, Adam and Mazi{\`e}res, David and Kozyrakis, Christos},
 title={{Shinjuku: Preemptive Scheduling for Microsecond-scale Tail Latency}},
 booktitle={16th USENIX Symposium on Networked Systems Design and Implementation (NSDI 19)},
 year={2019}, pages={345--360}, publisher={USENIX Association},
 note={\href{https://www.usenix.org/conference/nsdi19/presentation/kaffes}{Publisher version}}
}

@inproceedings{steere1999feedback,
 author={Steere, David C. and Goel, Ashvin and Gruenberg, Joshua and McNamee, Dylan and Pu, Calton and Walpole, Jonathan},
 title={{A Feedback-Driven Proportion Allocator for Real-Rate Scheduling}},
 booktitle={3rd Symposium on Operating Systems Design and Implementation (OSDI 99)},
 year={1999}, pages={145--158}, publisher={USENIX Association},
 note={\href{https://www.usenix.org/conference/osdi-99/feedback-driven-proportion-allocator-real-rate-scheduling}{Publisher version}}
}

@inproceedings{zheng2004swap,
 author={Zheng, Haoqiang and Nieh, Jason},
 title={{SWAP: A Scheduler with Automatic Process Dependency Detection}},
 booktitle={First Symposium on Networked Systems Design and Implementation (NSDI 04)},
 year={2004}, publisher={USENIX Association},
 note={\href{https://www.usenix.org/conference/nsdi-04/swap-scheduler-automatic-process-dependency-detection}{Publisher version}}
}

@inproceedings{palix2011faults,
 author={Palix, Nicolas and Thomas, Ga{\"e}l and Saha, Suman and Calv{\`e}s, Christophe and Lawall, Julia L. and Muller, Gilles},
 title={{Faults in Linux: Ten Years Later}},
 booktitle={Proceedings of the Sixteenth International Conference on Architectural Support for Programming Languages and Operating Systems (ASPLOS '11)},
 year={2011}, pages={305--318}, publisher={Association for Computing Machinery},
 doi={10.1145/1950365.1950401}, note={\href{https://doi.org/10.1145/1950365.1950401}{Publisher version}}
}

@inproceedings{engler2000checker,
 author={Engler, Dawson R. and Chelf, Benjamin and Chou, Andy and Hallem, Seth},
 title={{Checking System Rules Using System-Specific, Programmer-Written Compiler Extensions}},
 booktitle={4th Symposium on Operating Systems Design and Implementation (OSDI 00)},
 year={2000}, pages={1--16}, publisher={USENIX Association},
 note={\href{https://www.usenix.org/conference/osdi-2000/checking-system-rules-using-system-specific-programmer-written-compiler}{Publisher version}}
}

@inproceedings{klein2009sel4,
 author={Klein, Gerwin and Elphinstone, Kevin and Heiser, Gernot and Andronick, June and Cock, David and Derrin, Philip and Elkaduwe, Dhammika and Engelhardt, Kai and Kolanski, Rafal and Norrish, Michael and Sewell, Thomas and Tuch, Harvey and Winwood, Simon},
 title={{seL4: Formal Verification of an OS Kernel}},
 booktitle={Proceedings of the ACM SIGOPS 22nd Symposium on Operating Systems Principles (SOSP '09)},
 year={2009}, pages={207--220}, publisher={Association for Computing Machinery},
 doi={10.1145/1629575.1629596}, note={\href{https://doi.org/10.1145/1629575.1629596}{Publisher version}}
}

@inproceedings{gu2016certikos,
 author={Gu, Ronghui and Shao, Zhong and Chen, Hao and Wu, Xiongnan and Kim, Jieung and Sj{\"o}berg, Vilhelm and Costanzo, David},
 title={{CertiKOS: An Extensible Architecture for Building Certified Concurrent OS Kernels}},
 booktitle={12th USENIX Symposium on Operating Systems Design and Implementation (OSDI 16)},
 year={2016}, pages={653--669}, publisher={USENIX Association},
 note={\href{https://www.usenix.org/conference/osdi16/technical-sessions/presentation/gu}{Publisher version}}
}

@inproceedings{coccinelle2008,
 author={Padioleau, Yoann and Lawall, Julia L. and Hansen, Ren{\'e} Rydhof and Muller, Gilles},
 title={{Documenting and Automating Collateral Evolutions in Linux Device Drivers}},
 booktitle={Proceedings of the 3rd ACM SIGOPS/EuroSys European Conference on Computer Systems (EuroSys '08)},
 year={2008}, pages={247--260}, doi={10.1145/1352592.1352618},
 note={\href{https://doi.org/10.1145/1352592.1352618}{Publisher version}}
}

@inproceedings{tamches1999kerninst,
 author={Miller, Barton P. and Tamches, Ariel},
 title={{Fine-Grained Dynamic Instrumentation of Commodity Operating System Kernels}},
 booktitle={3rd Symposium on Operating Systems Design and Implementation (OSDI 99)},
 year={1999}, pages={117--130}, publisher={USENIX Association},
 note={\href{https://www.usenix.org/conference/osdi-99/fine-grained-dynamic-instrumentation-commodity-operating-system-kernels}{Publisher version}}
}

@inproceedings{cantrill2004dtrace,
 author={Cantrill, Bryan M. and Shapiro, Michael W. and Leventhal, Adam H.},
 title={{Dynamic Instrumentation of Production Systems}},
 booktitle={2004 USENIX Annual Technical Conference (USENIX ATC 04)},
 year={2004}, pages={15--28}, publisher={USENIX Association},
 note={\href{https://www.usenix.org/conference/2004-usenix-annual-technical-conference/dynamic-instrumentation-production-systems}{Publisher version}}
}

@inproceedings{yaseen2020aragog,
 author={Yaseen, Nofel and Arzani, Behnaz and Beckett, Ryan and Ciraci, Selim and Liu, Vincent},
 title={{Aragog: Scalable Runtime Verification of Shardable Networked Systems}},
 booktitle={14th USENIX Symposium on Operating Systems Design and Implementation (OSDI 20)},
 year={2020}, pages={701--718}, publisher={USENIX Association},
 note={\href{https://www.usenix.org/conference/osdi20/presentation/yaseen}{Publisher version}}
}

@misc{idle_source_snapshot,
 author={{Linux Kernel Contributors}},
 title={{Linux scheduler source at the idle-time validity correction}},
 year={2026}, month=sep,
 howpublished={Linux mainline source snapshot},
 note={Revision c6dcd97c8be75f052a1ca52cf79b03e7292962f1; kernel/sched/core.c, fair.c, idle.c, and sched.h. \href{https://kernel.googlesource.com/pub/scm/linux/kernel/git/torvalds/linux/+/c6dcd97c8be75f052a1ca52cf79b03e7292962f1/kernel/sched/}{Source snapshot}}
}

@misc{idle_guard_followup,
 author={Kaushik, Shubhang},
 title={{Re: [PATCH v3] sched/core: Skip rq$\rightarrow$avg\_idle update without a valid idle\_stamp}},
 year={2026}, month=aug,
 howpublished={Linux Kernel Mailing List. \href{https://lore.kernel.org/all/2bef9a46-0f37-c4d1-f1d3-b69639dc0ebe@gentwo.org/}{Source record}},
 note={August 17; repeated hackbench thread and process cases}
}
\endgroup
\end{document}